\documentclass[conference]{IEEEtran}
\usepackage{cite}
\usepackage{amsmath,amssymb,amsfonts, yhmath}
\usepackage{amsthm}
\usepackage{graphicx}
\usepackage{booktabs}
\usepackage{textcomp}
\usepackage{xcolor}
\usepackage{bbm}
\usepackage{algorithm}
\usepackage[noend]{algpseudocode}
\usepackage{framed,multirow}
\usepackage{mwe}
\usepackage{diagbox}
\usepackage{mathtools}

\usepackage[caption=false,font=normalsize,labelfont=sf,textfont=sf]{subfig}
\allowdisplaybreaks

\def\BibTeX{{\rm \mathcal{B}\kern-.05em{\sc i\kern-.025em b}\kern-.08em
    T\kern-.1667em\lower.7ex\hbox{\mathcal{E}}\kern-.125emX}}

\theoremstyle{plain}

\theoremstyle{remark}

\algrenewcommand\algorithmicrequire{\textbf{Input:}}
\algrenewcommand\algorithmicensure{\textbf{Output:}}
\usepackage[switch,columnwise]{lineno}
\DeclareRobustCommand{\IEEEauthorrefmark}[1]{\smash{\textsuperscript{\footnotesize #1}}}
\IEEEoverridecommandlockouts

\allowdisplaybreaks[4]

\begin{document}

% \title{Conference Paper Title*\\
% {\footnotesize \textsuperscript{*}Note: Sub-titles are not captured in Xplore and
% should not be used}
% \thanks{Identify applicable funding agency here. If none, delete this.}
% }
\title{
WMAS: A Multi-Agent System Towards Intelligent and Customized Wireless Networks
\thanks{The first two authors contributed equally to this work and $\dag$ marked the corresponding author. This work was supported in part by the National Natural Science Foundation of China under Grant 12271289.  }
}\vspace{-0.2em}
\author{\IEEEauthorblockN{
Jingchen Peng\IEEEauthorrefmark{1}, 
Dingli Yuan\IEEEauthorrefmark{1}, 
Boxiang Ren\IEEEauthorrefmark{1},
Jie Fan\IEEEauthorrefmark{1},
Hao Wu\IEEEauthorrefmark{1},
and Lu Yang\IEEEauthorrefmark{2}$^\dag$}
\IEEEauthorblockA{\IEEEauthorrefmark{1}Department of Mathematical Sciences, Tsinghua University, Beijing, China}
\IEEEauthorblockA{\IEEEauthorrefmark{2}Wireless Technology Lab, Central Research Institute, 2012 Labs, Huawei Tech. Co. Ltd., China}
  \IEEEauthorblockA{Email: yanglu87@huawei.com}\vspace{-1.2em}

 }

% \author{\IEEEauthorblockN{Boxiang Ren\IEEEauthorrefmark{1}, Jingchen Peng\IEEEauthorrefmark{1}, Han Hao\IEEEauthorrefmark{1}, Chaowen Deng\IEEEauthorrefmark{1}, Junyuan Wang\IEEEauthorrefmark{2}$^\dag$ and Hao Wu\IEEEauthorrefmark{1}}

%   \IEEEauthorblockA{\IEEEauthorrefmark{1}Department of Mathematical Sciences, Tsinghua University, Beijing, China}
%   \IEEEauthorblockA{\IEEEauthorrefmark{2}College of Electronic and Information Engineering, Tongji University, Shanghai, China}
%   \IEEEauthorblockA{Email: junyuanwang@tongji.edu.cn}
%  }

\maketitle

\begin{abstract}

% %  原
% Recent advancements in Artificial Intelligence (AI) agents have demonstrated remarkable performance, making agent-driven systems a promising solution for more intelligent and adaptive wireless networks. 
% In this paper, we propose a Wireless Multi-Agent System (WMAS), which can provide intelligent and customized services for different user terminals (UEs).
% v1  --------------
The fast development of Artificial Intelligence~(AI) agents provides a promising way for the realization of intelligent and customized wireless networks. 
In this paper, we propose a Wireless Multi-Agent System~(WMAS), which can provide intelligent and customized services for different user equipment~(UEs).
% --------------
Note that orchestrating multiple agents carries the risk of malfunction, and multi-agent conversations may fall into infinite loops.
It is thus crucial to design a conversation topology for WMAS that enables agents to complete UE task requests with high accuracy and low conversation overhead.  
To address this issue, we model the multi-agent conversation topology as a directed acyclic graph and propose a reinforcement learning-based algorithm to optimize the adjacency matrix of this graph.
As such, WMAS is capable of generating and self-optimizing multi-agent conversation topologies, enabling agents to effectively and collaboratively handle a variety of task requests from UEs.
Simulation results across various task types demonstrate that WMAS can achieve higher task performance and lower conversation overhead compared to existing multi-agent systems. These results validate the potential of WMAS to enhance the intelligence of future wireless networks.
\end{abstract}


\section{Introduction}

Conventional wireless networks primarily aim to provide communication services to users. However, with the swift progression of artificial intelligence (AI) technologies, especially large language models (LLMs) and AI agents, the vision for sixth-generation (6G) networks goes beyond just connectivity. In particular, 6G is anticipated to integrate AI, computing, and sensing alongside connectivity, broadening its scope to include new resources, features, and design paradigms. This evolution calls for a more intelligent wireless network that can offer customized services to users autonomously~\cite{jiang2024large,xu2024large,saad2025artificial}.

AI agents, with LLMs as their cognitive core, can intelligently observe the environment, make decisions, and take actions without predefined instructions or direct human intervention~\cite{wang2024survey,tong2024wirelessagent}. 
In addition, unlike traditional network entities that rely on symbolic languages, AI agents communicate via natural language, enabling more flexible and human-like interactions. This allows them to deliver a broad range of services, including data transmission, voice calls, multimedia applications, sensing tasks, and AI-based services \cite{letaief2021edge,wen2023task,qian2024chatdev}. 
Therefore, AI agents can serve as a compelling enabler for future wireless networks. \par

Despite the aforementioned potential, single-agent systems exhibit notable limitations. Relying on self-dialogue, single agents lack cognitive diversity, resulting in repetitive reasoning and degraded performance~\cite{subramaniam2025multiagent}. Furthermore, a single agent is often insufficient for handling complex or multi-step tasks~\cite{liang2024encouraging}. 
In contrast, Multi-Agent Systems (MASs), equipped with role specialization and collaborative conversation, can benefit from diversity gains, so that a MAS can deal with more complicated tasks and achieve better performance~\cite{li2023camelcommunicativeagentsmind,wu2024autogen,qian2025scaling}.
Unfortunately, MASs still have the risk of malfunction, and multi-agent conversations may loop infinitely,  which will lead to substantial conversation overhead~\cite{wang2024reasoning}. 
% This challenge is particularly pertinent across the diverse deployment scenarios of MAS for future wireless networks, whether agents are hosted in centralized settings (e.g., access or core networks) or distributed network devices (e.g., base stations~(BS), Multi-access Edge Computing~(MEC) servers, or user terminals).
Therefore, it is significant to design a collaboration scheme to ensure that the MAS can solve various tasks with superior performance but low conversation overhead. \par
One way to model the collaboration scheme of a MAS is the conversation topology, since it reflects the temporal and logical dependencies inherent in agents’ dialogue.
Currently, most of the MASs are operated based on a predefined conversation topology, such as CAMEL~\cite{li2023camelcommunicativeagentsmind}, AutoGen~\cite{wu2024autogen}, MacNet~\cite{qian2025scaling}, etc. These fixed designs require substantial manual engineering, and lack the room for reducing the high redundancy of multi-round conversations. To address these issues, GPTSwarm~\cite{zhuge2024gptswarm} was proposed, which utilized an optimizable graph to model the multi-agent conversations, and leveraged the reinforcement learning~(RL) to achieve the topology's self-improvement.
However, this scheme is only feasible for a single round of dialogue, and is not suitable for multi-round conversations. Moreover, it focuses solely on the task performance of the MAS, without accounting for conversation overhead. AgentPrune~\cite{zhang2025cut} introduced a spatial-temporal graph to represent the conversation topology, and proposed a pruning approach to reduce the conversation overhead. Nevertheless, such pruning strategy keeps fixed for all the dialogue rounds, ignoring the temporal variation. In addition, its one-shot pruning mechanism compromised the generality and adaptability of the MAS. \par
In this paper, we propose a Wireless Multi-Agent System~(WMAS) to enable intelligent and customized wireless networks. 
Through multi-agent conversations, WMAS can provide customized services for different user requests, especially for future AI-based requests.
% WMAS supports multi-round agent dialogues, enhancing network intelligence and adaptability while effectively utilizing edge resources to provide customized, low-latency services.
% A key factor influencing the performance of multi-agent systems is the underlying conversation topology, as it captures the temporal and logical dependencies across dialogue rounds, which in turn shape the system’s coordination mechanism.
Considering the critical role of conversation topology in MAS, we first model it as a Directed Acyclic Graph (DAG), referred to as the conversation graph, which serves as our optimization object.
Second, we propose a metric $\mathcal{C}_{\text{WMAS}}$ that jointly considers the task performance and the conversation overhead as the objective function. 
Furthermore, given the combinatorial complexity of searching over graph structures, we propose a probabilistic heatmap, in which each entry represents the likelihood and importance of a potential conversation edge between agents. 
This heatmap acts as a continuous relaxation of the adjacency matrix of the conversation graph. 
% which can convert the original discrete objective function into a continuous one.   
The gradient of our objective function can thus be approximated using policy gradient methods from RL. 
Fourth, we propose an optimization algorithm to improve the heatmap iteratively. 
As a result, our WMAS can improve the conversation topology by itself.
Simulation results on general reasoning, mathematical reasoning, and code generation demonstrate that WMAS outperforms existing MASs in terms of the performance of task execution as well as the conversation overhead.

\section{System Model}
In this section, we introduce our WMAS, including its framework and mathematical representations, which dynamically orchestrate multi-agent collaboration, and provide customized services for different users.

\subsection{Design of the Wireless Multi-Agent System}
% \begin{figure}[htbp]
%     \centering
% \includegraphics[width=1.0\linewidth]{figure/system_model_8.png}
%     \caption{System model of the Multi-Agent Conversation System (MACS).}
%     \label{fig: system-model}
% \end{figure}

\begin{figure}[t]
    \centering
\includegraphics[width=0.95\linewidth]{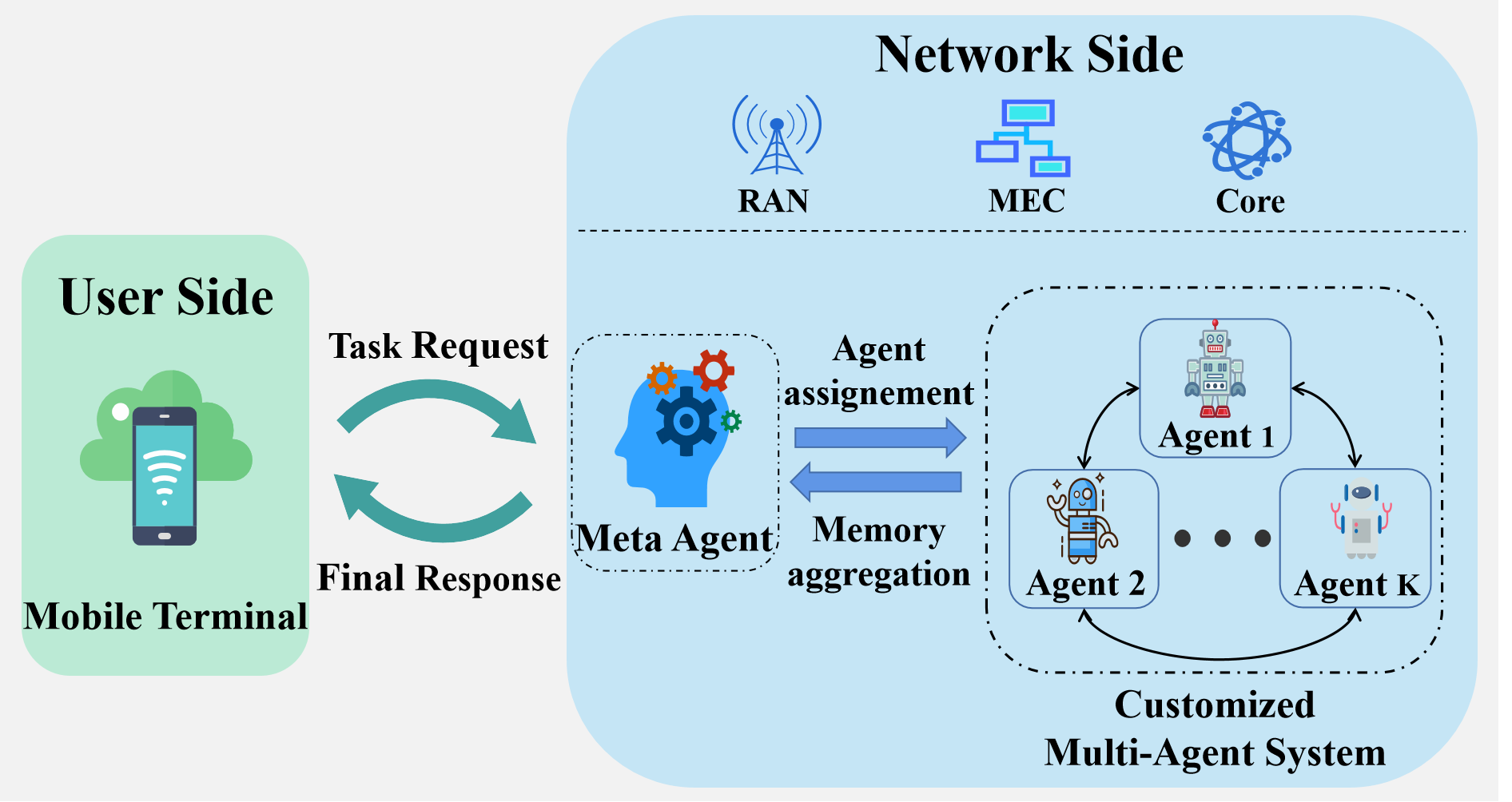}
    \caption{The overall framework of the Wireless Multi-Agent System (WMAS)}
    \label{fig: system-model}
\end{figure}

As illustrated in Fig.~\ref{fig: system-model}, WMAS consists of two core components: a meta agent and a customized MAS. 
The meta agent is responsible for understanding the user's task request, orchestrating a customized MAS to deal with this task, and synthesizing the final result based on the MAS’s conversation history. 
The customized MAS is responsible for providing a solution scheme for a given task through multi-round conversations. 
WMAS has self-optimization capabilities. During task execution, the meta agent implements an optimization approach for the conversation topology, and samples an improved topology. The MAS then executes subsequent tasks based on this topology.

% WMAS has self-optimization capabilities. During task execution, the meta agent, based on feedback from previous task results, performs the conversation topology optimization algorithm. It then samples an improved topology for the MAS, which executes subsequent tasks according to this topology.
\par

WMAS can be deployed at the wireless network, with a centralized way or a distributed way\footnote{The centralized way means the whole WMAS framework is deployed at a single network entity, such as the core network (CN), or the radio access network (RAN), or the mobile edge cloud (MEC), etc. The distributed way refers to the components of WMAS deployed at different network entities.}. 
In this paper, we consider the centralized way. When the wireless network receives a task request from the user side, the meta agent at the network side should understand this task request, classify it, and orchestrate a customized MAS to tackle it.
The meta agent determines the number of participating agents $K$, the role of each agent, the number of dialogue rounds $T$, and an initial conversation topology for these agents. For example, if the network side receives a code generation task, the meta agent may choose three agents, with the roles of Algorithm Designer, Programming Expert, and Test Analyst, respectively~\cite{dong2024self}.
In addition, these three agents will receive an initial conversation topology from the meta agent, so that they can collaborate with each other according to the corresponding workflow. 
Note that this conversation topology can be optimized by WMAS itself, based on our RL-based approach. All the conversation history will be stored in the memory of the meta-agent. After $T$ rounds of interaction, the meta agent will analyze the entire dialogue and produce the final response to the mobile terminal.

\subsection{Mathematical Definition of a Single Agent and a MAS}
An agent in our WMAS can be mathematically represented as $a =\{M, r, c, m, t, k \}$~\cite{tran2025multi}, where: 
\begin{itemize}
 \item \textbf{Model} $M$: The large language model (LLM), embedded in the agent, is responsible for understanding the natural language inputs, including intent, context, semantics, etc., and generating responses to the input. WMAS is compatible with various LLMs, such as OpenAI’s GPT, Meta’s LLaMA, Alibaba’s Qwen, etc.
 
\item \textbf{Role} $r$: The functional description of an agent. An agent knows its role through prompt engineering. MAS with a role-playing mechanism can improve its performance effectively.
\item \textbf{Communication} $c$: The communication module enables the agent to exchange information with other agents or external components. It supports standard networking protocols (e.g., TCP/IP, HTTP) and handles message transmission and reception.

\item \textbf{Memory} $m$: The internal memory module that stores intermediate outputs generated by the agent itself, as well as messages received from other agents. This component supports continuity across multi-round conversations and facilitates self-improvement over time.

 \item \textbf{Tool} $t$: The external tool that an agent can invoke to perform functions beyond its internal model, such as code execution, web search, or simulation of custom communication models (e.g., wireless channel models).

 \item \textbf{Knowledge Base} $k$: The repository of domain-specific knowledge, including communication standards, technical documents, and research papers, which provides background information for solving complex tasks in wireless communication.
 \end{itemize} \par
Based on the single-agent structure, we further define the MAS as a tuple \(\mathcal{S} = (\mathcal{A}, \mathcal{O},\mathcal{C}, \mathcal{F})\):
\begin{itemize}
    \item \(\mathcal{A} = \{a_k\}_{k=1}^K\): A set of \(K\) agents, each represented as \(a_i = \{M_i, r_i, c_i, m_i, t_i, k_i\}\). The number of agents \(K\) is determined according to the task type, and can be dynamically adjusted based on system requirements.
    \item \(\mathcal{O}\): The overall task objective of the MAS, representing the goal that the agents aim to achieve collaboratively.
     \item \(\mathcal{C}\): The conversation topology, which defines the directional dependencies among agents during multi-round conversations. It determines the collaboration scheme of the MAS, specifying which agent sends information to which others and in what order.
\item \(\mathcal{F}\): The workflow that governs agent execution order. Built on \(\mathcal{C}\), it enforces a well-defined execution order to avoid circular dependencies.
     
%      \item \(\mathcal{F}\): The workflow that governs agent execution order. Built on \(\mathcal{C}\), it ensures each agent activates only after receiving all required inputs, thereby avoiding communication stalls and deadlocks.%
% \footnote{A communication stall occurs when an agent waits indefinitely for input from others; a deadlock arises from circular dependencies among agents.}
\end{itemize}\par
During execution, each agent is activated following the workflow \(\mathcal{F}\). Upon activation, agent \(a_i\) receives input \(x_i\) from its upstream agents as specified by the conversation topology \(\mathcal{C}\). It then generates a response \(y_i\) by selecting the most probable output based on its internal state, the task objective \(\mathcal{O}\), and the input \(x_i\):
\[
y_i^* = \arg\max_{y_i} P(y_i \mid a_i, x_i, \mathcal{O}).
\]
The response \(y_i^*\) is then forwarded to subsequent agents according to \(\mathcal{C}\), continuing the multi-round conversation.

\section{Self-Optimization of Conversation Topology} \label{Sec: our Method}
In MAS, the conversation topology $\mathcal{C}$ is essential, as it represents the temporal and logical dependencies of agent dialogues, and can dominate the performance of a MAS. However, optimizing the conversation topology is challenging due to the difficulty in finding a suitable way to model the agents' dialogues mathematically. 
In this section, we model the conversation topology as a conversation graph, motivated by its structural similarity to a DAG. We then propose a performance metric to evaluate the efficiency of the conversation graph, which serves as our objective function.\par
Nevertheless, it is inherently intractable to optimize the conversation graph due to its combinatorial complexity and discrete structure.
To address this issue, we introduce a probabilistic heatmap as a continuous relaxation and develop a heatmap optimization algorithm that uses policy gradient methods to iteratively refine it.

\subsection{Graph Modeling and Metric Design}\label{sec: graph-model}

\begin{figure}[t]
    \centering
    \includegraphics[width=0.95\linewidth]{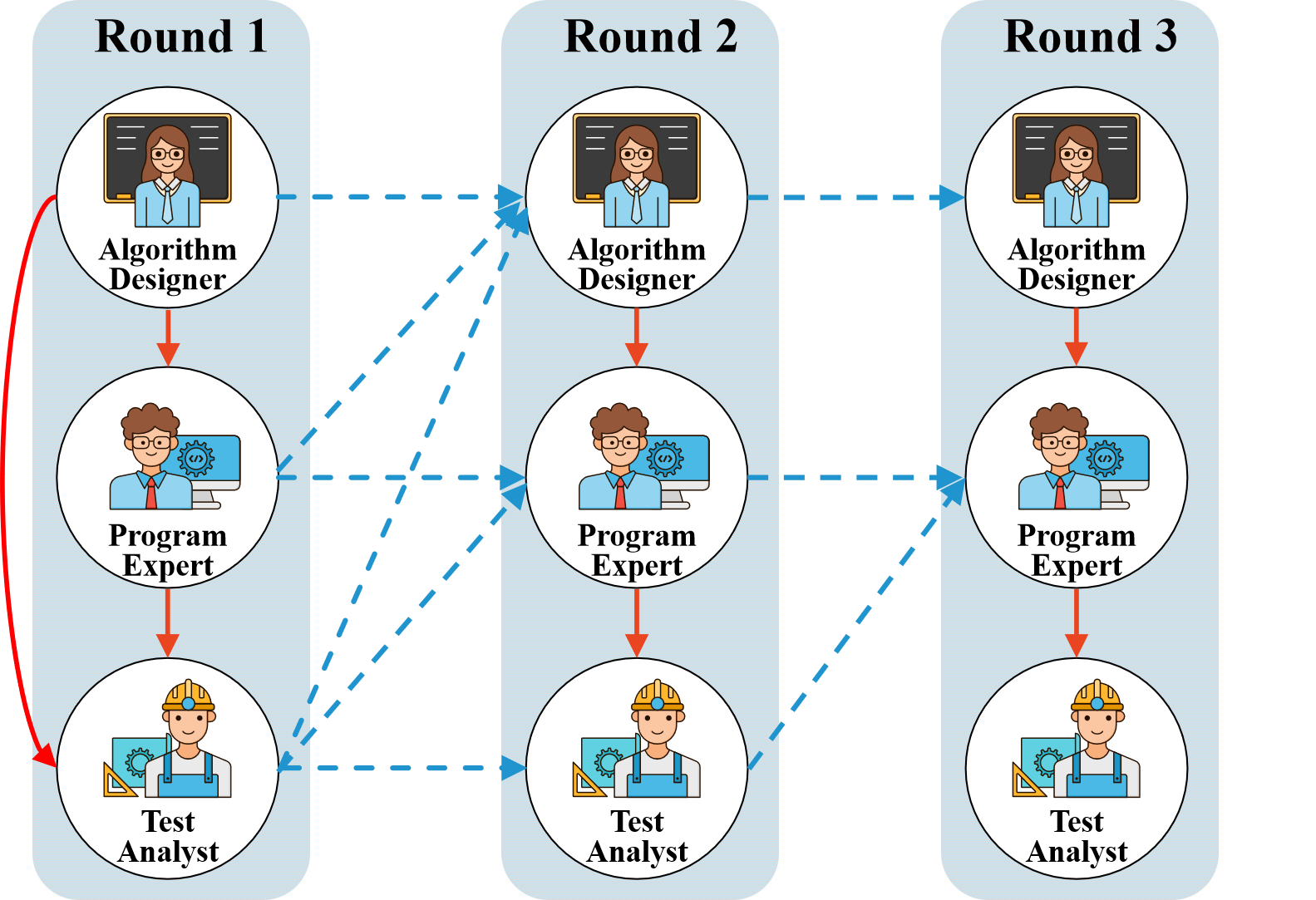}
    \caption{A schematic illustration of the conversation graph for a code generation task involving three rounds ($T=3$) and three agents ($K=3$). Solid red arrows denote information flow among agents within the same round, whereas dashed blue arrows signify information transfer across different rounds.}
    \label{fig:multi-role}
\end{figure}

We first model the conversation topology $\mathcal{C}$, which involves \(K\) agents over \(T\) dialogue rounds, as a directed conversation graph \(\mathcal{G} = (\mathcal{V}, \mathcal{E})\). 
To ensure correct execution order and prevent infinite loops, $\mathcal{G}$ is required to be a DAG so that a valid workflow can be derived via topological ordering of agent nodes~\cite{bondy1976graph}. A schematic illustration of the conversation graph is shown in Fig.~\ref{fig:multi-role} \par
The structure of $\mathcal{G}$ is defined as follows. Each agent \(a_k^{(t)}\), representing the \(k\)-th agent in round \(t\), is mapped to a unique node \(v_i \in \mathcal{V}\), where \(i = (t - 1)K + k\). This mapping yields a total of \(|\mathcal{V}| = TK\) nodes, with each node corresponding to a specific agent instance at a specific round. 

A directed edge \(e_{ij} = (v_i, v_j) \in \mathcal{E}\) denotes a conversation link from \(v_i\) to \(v_j\), meaning that the output of agent \(v_i\) serves as input to agent \(v_j\). The overall connectivity of \(\mathcal{G}\) can be represented by an adjacency matrix \(\mathbf{A} \in \{0,1\}^{TK \times TK}\), where \(\mathbf{A}[i,j] = 1\) if \(e_{ij} \in \mathcal{E}\), and \(\mathbf{A}[i,j] = 0\) otherwise.\par
To reflect practical multi-agent conversation scenarios, structural constraints are imposed on \(\mathbf{A}\) to ensure temporal and logical consistency by eliminating infeasible edges. These constraints include the following:
\begin{itemize}
      \item \(\mathbf{A}[i,i] = 0\). Self-loops are disallowed; an agent cannot engage in dialogue with itself within the same round.
      \item \(\mathbf{A}[i,j] = 0\) if agent \(v_i\) belongs to a later round than agent \(v_j\), i.e., \(\lfloor \frac{i}{K} \rfloor > \lfloor \frac{j}{K} \rfloor\), where \(\lfloor \cdot \rfloor\) denotes the largest integer less than or equal to the value. This prevents information from flowing backward across dialogue rounds.
        \item \(\mathbf{A}[i,j] = 0\) if agent \(v_j\) belongs to a round more than one step after \(v_i\), i.e., \(\lfloor \frac{j}{K} \rfloor - \lfloor \frac{i}{K} \rfloor > 1\). This ensures messages are only passed to agents in the same or the next round.
\end{itemize}
These constraints ensure causality and preserve the sequential structure of multi-round conversation.
We next focus on identifying which conversation graph leads to the optimal multi-agent collaboration, and thus propose a novel evaluation metric: 
\begin{equation}\label{eq: 1}
    \mathcal{C}_{\text{WMAS}}(\mathcal{G}) = \boldsymbol{\phi}(\mathcal{G}, D) - \beta \|\mathbf{A}\|_1.
\end{equation}
This metric is composed of two complementary components. The utility function $\boldsymbol{\phi}(\mathcal{G}, D)$ measures the task performance of WMAS on dataset $D$ under the graph $\mathcal{G}$. The second term, $\|\mathbf{A}\|_1$, is the $L_1$-norm of the adjacency matrix, which corresponds to the total number of conversation edges in the graph and serves as an indicator of conversation overhead. The hyperparameter $\beta$ controls the trade-off between task quality and conversation overhead.\par
A higher value of $\mathcal{C}_{\text{WMAS}}(\mathcal{G})$ indicates better overall performance, balancing effectiveness and efficiency. Our objective is thus to find the optimal conversation graph $\mathcal{G}$ by solving:
\begin{equation}\label{eq: 2}
    \begin{aligned}
\max_{\mathcal{G}}\quad &\mathcal{C}_{\text{WMAS}}(\mathcal{G}) \\
\text{s.t. }\quad &\mathcal{G} \text{ is a DAG}
\end{aligned}
\end{equation}

\subsection{Continuous Relaxaion and Heatmap Optimization}

\begin{algorithm}[t]
\caption{DAG Sampling from Probabilistic Heatmap \(\mathbf{H}\)}
\label{alg: sampling}
\begin{algorithmic}[1]
\State \textbf{Input:} Heatmap \(\mathbf{H} \in [0,1]^{TK \times TK}\)
\State \textbf{Output:} Conversation graph \(\mathcal{G} = (\mathcal{V}, \mathcal{E})\), log-probability \(\log p_{\mathbf{H}}(\mathcal{G})\)
\State Initialize \(\mathcal{E} \leftarrow \emptyset\), \(\mathcal{V} = \{v_1, v_2, \dots, v_{TK}\}\), \(\log p \leftarrow 0\)
\For{each pair \((i, j)\) where \(i \ne j\)}
    \If{adding \((v_i, v_j)\) to \(\mathcal{E}\) does not create a cycle}
        \State Sample \(e_{ij} \sim \text{Bernoulli}(\mathbf{H}[i,j])\)
        \If{\(e_{ij} = 1\)}
            \State Add \((v_i, v_j)\) to \(\mathcal{E}\)
            \State \(\log p \leftarrow \log p + \log(\mathbf{H}[i,j])\)
        \Else
            \State \(\log p \leftarrow \log p + \log(1 - \mathbf{H}[i,j])\)
        \EndIf
    \EndIf
\EndFor
\State \Return \(\mathcal{G} = (\mathcal{V}, \mathcal{E}), \log p_{\mathbf{H}}(\mathcal{G}) = \log p\)
\end{algorithmic}
\end{algorithm}

% Optimizing the conversation graph is inherently challenging due to the combinatorial explosion of possible configurations and the discrete nature of the adjacency matrix $\mathbf{A}$, which limits the applicability of gradient-based methods.

As a relaxation of the adjacency matrix \(\mathbf{A}\), we introduce a probabilistic heatmap \(\mathbf{H} \in [0,1]^{TK \times TK}\), where each entry replaces a binary variable in \(\mathbf{A}\) with a continuous one. \par
Each entry \(\mathbf{H}[i,j]\) represents the probability that edge \(e_{ij}\) exists in the graph, while also reflecting its importance for task execution. A sigmoid function is applied to ensure that the value lies within the range \((0,1)\). The heatmap $\mathbf{H}$ inherits the structural constraints imposed on $\mathbf{A}$. Specifically, entries corresponding to infeasible edges (as defined in Section~\ref{sec: graph-model}) are zero to follow the constraints. The set of all structurally valid heatmaps is denoted by $\mathbb{H}$.
\par
With the probabilistic heatmap \(\mathbf{H}\) defined, we now relax the original discrete optimization objective~\eqref{eq: 2} into the following continuous optimization problem:
\begin{equation}\label{eq: objective}
 \max_{\mathbf{H} \in \mathbb{H}} \; \mathbb{E}_{\mathcal{G} \sim \mathbf{H}} \left[\boldsymbol{\phi}(\mathcal{G}, D) - \beta \|\mathbf{A}\|_1\right]
\end{equation}
Here, $\mathcal{G} \sim \mathbf{H}$ denotes a graph that follows the probability distribution parameterized by $\mathbf{H}$, under the DAG constraint.\par
% A sampling-based method is employed to approximate this distribution by constructing a DAG-form conversation graph $\mathcal{G}$ based on $\mathbf{H}
% $. 
A sampling-based method is employed to approximate this distribution.
The sampling algorithm iterates over all feasible directed edges and includes each edge $(v_i, v_j)$ with probability $\mathbf{H}[i,j]$, as long as its inclusion does not create a cycle. 
The log-probability 
$\log p_{\mathbf{H}}(\mathcal{G})$ of $\mathcal{G}$ is also computed by summing the log-likelihoods of both included and excluded edges. The complete sampling process is described in Algorithm~\ref{alg: sampling}.
\par
The utility function in~\eqref{eq: objective} is non-differentiable and intractable, as it typically relies on external Application Programming Interfaces (APIs) or compilers \cite{li-etal-2023-api,hendrycks2021measuring}. We therefore apply policy gradient methods~\cite{williams1992simple} to obtain an unbiased gradient estimation~\cite{zhuge2024gptswarm}:
\begin{equation}
    \nabla_{\mathbf{H}}\ \mathbb{E}_{\mathcal{G} \sim \mathbf{H}} \left[ \boldsymbol{\phi}(\mathcal{G}, D)\right] \approx \frac{1}{N}\sum_{n=1}^N \boldsymbol{\phi}(\mathcal{G}_n,D) \nabla_{\mathbf{H}}\log p_{\mathbf{H}}(\mathcal{G}_n)
\end{equation}
Here, $\log p_{\mathbf{H}}(\mathcal{G}_n)$ denotes its log-probability of the $n$-th sampled graph, computed directly during the sampling process in Algorithm~\ref{alg: sampling}. Then the gradient approximation of \eqref{eq: objective} can be obtained as: 
\begin{equation}
\begin{aligned}
    &\nabla_{\mathbf{H}}\  \mathbb{E}_{\mathcal{G} \sim \mathbf{H}} \left[ \boldsymbol{\phi}(\mathcal{G}, D) - \beta \|\mathbf{H}\|_{1} \right] \\
    \approx &\frac{1}{N}\sum_{n=1}^N \boldsymbol{\phi}(\mathcal{G}_n,D) \nabla_{\mathbf{H}}\log p_{\mathbf{H}}(\mathcal{G}_n)-\beta  \partial_{\mathbf{H}}  \|\mathbf{H}\|_{1}.
\end{aligned}
\end{equation}
\par
Based on the above, we propose a heatmap optimization algorithm. The optimization procedure, implemented via a gradient descent variant, is outlined in Algorithm~\ref{alg: optimization}. 
% As \(\mathbf{H}\) is iteratively optimized, the quality of the graph sampled from it improves, leading to an enhanced conversation topology. 
This enables WMAS to self-optimize its multi-agent collaboration. Specifically, WMAS first optimizes the heatmap \(\mathbf{H}\), then samples the improved conversation topology from \(\mathbf{H}\), which guides agent collaboration during task execution. Through this iterative process, task effectiveness and efficiency of WMAS are progressively improved.

% Once $\mathbf{H}$ is determined, a high-quality conversation graph can be directly sampled from it to effectively guide task execution in the WMAS.

\begin{algorithm}[t]
\caption{Heatmap Optimization}
\label{alg: optimization}
\begin{algorithmic}[1]
\State \textbf{Input:} Dataset \(D\), learning rate \(\eta\), trade-off factor \(\beta\), initial heatmap \(\mathbf{H}\), number of iterations \(M\), sample size \(N\)
\State \textbf{Output:} Optimized heatmap \(\mathbf{H}\)

\For{\(m = 1\) \textbf{to} \(M\)}
    \State Sample mini-batch \(D' \subseteq D\)
\State Sample $\mathcal{G}_n$ from $\mathbf{H} $ for \( n = 1, 2, \dots, N \) \hfill (Algorithm~\ref{alg: sampling})
\State Approximate gradient
    \[
    \nabla \mathbf{H} \leftarrow \frac{1}{N} \sum_{n=1}^N \boldsymbol{\phi}(\mathcal{G}_n, D') \nabla_{\mathbf{H}} \log p_{\mathbf{H}}(\mathcal{G}_n) - \beta \partial_{\mathbf{H}} \|\mathbf{H}\|_1
    \]
\State Update heatmap:
  $\mathbf{H} \leftarrow \text{sigmoid}\left( \mathbf{H} + \eta \nabla \mathbf{H} \right)$
\State Project $\mathbf{H}$ onto $\mathbb{H}$
\EndFor

\State \Return \(\mathbf{H}\)
\end{algorithmic}
\end{algorithm}

\section{Simulation Results and Analysis}
\begin{figure*}[t]
    \centering
    \includegraphics[width=0.95\linewidth]{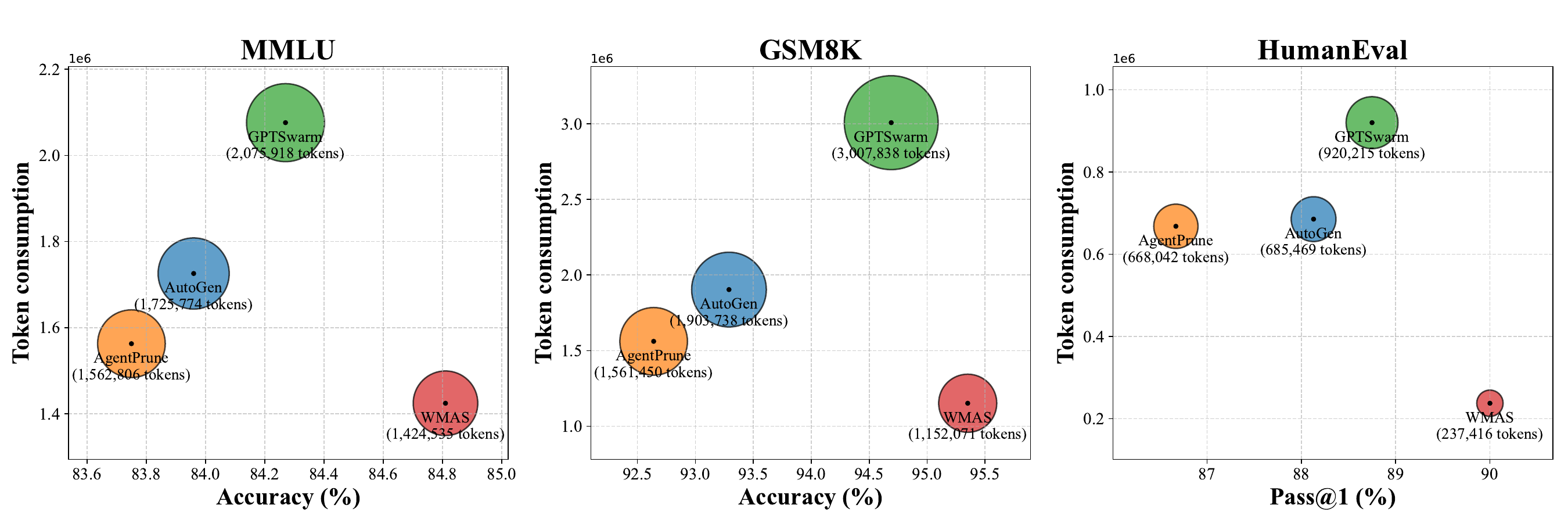}
    \caption{Task performance and conversation overhead tested from three datasets: MMLU, GSM8K, and HumanEval. The area of each bubble is proportional to the token consumption.}
    \label{fig:comm_cost_multiagent}
    \vspace{-1em}
\end{figure*}

\addtolength{\topmargin}{0.06in}
\subsection{Simulation Settings}
\subsubsection{Datasets} In our experiments, we evaluate the performance of WMAS on three types of tasks. First, the general reasoning task with the MMLU dataset~\cite{hendrycks2021measuring}, which contains multiple choice questions from academic subjects at the high school and college levels. Second, mathematical reasoning using GSM8K~\cite{cobbe2021trainingverifierssolvemath}, a high-quality dataset of math word problems. Third, code generation, using the most classic dataset HumanEval~\cite{chen2021evaluating}.
\subsubsection{Baselines} 
We compare WMAS with both single-agent frameworks and existing MAS frameworks. For single-agent frameworks, we consider direct inference~(referred to as Vanilla) and chain-of-thought prompting~(CoT)~\cite{wei2022chain} as the baselines. For existing MAS frameworks, we choose AutoGen~\cite{wu2024autogen}, GPTSwarm~\cite{zhuge2024gptswarm}, and AgentPrune~\cite{zhang2025cut} as the benchmarks.
\subsubsection{Implementation Details} 
In our experiments, all agents are deployed with the LLM of Qwen2.5-72B-Instruct model\footnote{All kinds of LLMs are supportive by WMAS. Here we choose an open-source LLM as an example.} via the Qwen API~\cite{yang2024qwen2}. For all baselines, we adopt their publicly released code and default configurations. 
% For WMAS, following~\cite{zhang2025cut}, the number of dialogue rounds is set to \(T = 2\) for general and mathematical reasoning tasks, and \(T = 4\) for code generation. 
For WMAS, the number of optimization iterations is set to \(15\), with a sample size of 4 per iteration. The learning rate \(\eta\) and trade-off parameter \(\beta\) are both set to \(1 \times 10^{-1}\). \par
In WMAS, we further customize agent roles for different tasks based on standard configurations~\cite{zhang2025cut,dong2024self}. For general reasoning, the roles include Knowledge Expert, Wiki Searcher, Critic, Mathematician, Programmer, Doctor, and Economist. For mathematical reasoning, the agents are Math Solver, Mathematical Analyst, Programming Expert, and Inspector. For code generation, we employ Algorithm Designer, Programming Expert, and Test Analyst. \par
\subsubsection{Metric}
We focus on two key aspects: task performance and conversation overhead.  
To measure the performance of a WMAS dealing with various tasks, we adopt a task-specific utility function \(\boldsymbol{\phi}\). For MMLU, \(\boldsymbol{\phi}\) corresponds to the answer accuracy. For GSM8K, it computes the accuracy via string matching between the predicted answer and the ground-truth. For HumanEval, \(\boldsymbol{\phi}\) is defined by pass@1~\cite{zhang2025cut}. To evaluate the conversation overhead, we choose the total token consumption, which includes both the tokens received by all agents and the tokens sent from all agents.

\subsection{Results and Discussion}
% \begin{table}[htbp]
%     \centering
%     \caption{Task performance of all methods on MMLU, GSM8K, and HumanEval. Metrics are task-specific: accuracy for MMLU and GSM8K, and pass@1 for HumanEval.}
%     \label{tab:performance_comparison}
%     \begin{tabular}{l|c|c|c}
%         \hline
%         \textbf{Method} & \textbf{MMLU} & \textbf{GSM8K} & \textbf{HumanEval} \\
%         \hline
%         Vanilla       &   82.35	& 91.02
% 	& 85.28

%           \\
%         CoT           &    83.66           &  92.19              &           86.67        \\
%         AutoGen       &    84.50          &      93.29        &        88.13          \\
%         GPTSwarm      &      84.96         &      94.69
%           &     88.75 
%             \\
%         AgentPrune    &   83.75          &      92.64     &            86.67        \\
%         \hline
%         \textbf{WMAS} &    \textbf{85.40}        &  \textbf{95.44}       & \textbf{90.00} \\
%         \hline
%     \end{tabular}
% \end{table}

We first compare the performance of WMAS with baseline methods on three datasets. As shown in Table~\ref{tab:performance_comparison}, WMAS consistently outperforms the best results of all baselines with performance gains ranging from 0.54\% to 1.25\%. 
Moreover, as the number of dialogue rounds $T$ increases, WMAS's performance are steadily improved on all datasets. These results verify that multi-round conversations can enhance agents' cognitive diversity and enable them to handle more complicated tasks collaboratively.
In addition, it can be observed that the MAS framework outperforms the single-agent framework~(Vanilla and CoT), underscoring the necessity of adopting multi-agent architectures for future wireless networks. \par
\setlength{\tabcolsep}{4pt}
   \begin{table}[b]
    \centering
    \caption{Task performance of agent frameworks on MMLU, GSM8K, and HumanEval. Bold values indicate results that outperform all baseline methods.}
    \label{tab:performance_comparison}
    \begin{tabular}{c|c|c|c|c}
        \hline
    \textbf{Framework}    &\textbf{Method} & \textbf{MMLU} & \textbf{GSM8K} & \textbf{HumanEval} \\
        \hline
\multirow{2}{*}{Single-agent} &      Vanilla       &   82.35	& 91.02
	& 85.28

          \\
    &    CoT           &    83.66           &  92.19              &           86.67        \\ \hline
 \multirow{3}{*}{\shortstack{Multi-agent\\ (Existing methods)}}     &  AutoGen       &    83.96          &      93.29        &        88.13          \\
           &     AgentPrune    &   83.75          &      92.64     &            86.67        \\
   &     GPTSwarm      &      84.27        &      94.69
          &     88.75 \\
        \hline
\multirow{3}{*}{\shortstack{Multi-agent\\(Ours)}} & WMAS~(T=1) 
            & 83.70 & 93.60 & 86.25 \\ % WMAS(T=1)
      &  WMAS~(T=2)     & \textbf{84.81} & \textbf{95.35} & 88.13 \\ % WMAS(T=2)
     & WMAS~(T=3)       & \textbf{84.96} & \textbf{95.44} & \textbf{90.00} \\ % WMAS(T=3)
        \hline
    \end{tabular}
\end{table}

% Moreover, we observe that as the number of dialogue rounds \(T\) increases, WMAS’s performance improves significantly. For MMLU and GSM8K, performance increases from \(T=1\) to \(T=2\) and then stabilizes as \(T\) moves from 2 to 3. For HumanEval, performance continues to improve as \(T\) increases from 1 to 3. These results confirm that multi-round conversation enhance agents' cognitive diversity, enabling them to handle more complex tasks.

We further investigate the conversation overhead of different MASs. For MMLU and GSM8K, since WMAS performs similarly when \(T\) equals 2 and 3, we set \(T=2\) to reduce conversation overhead. For HumanEval, we set \(T=3\). 
Fig.~\ref{fig:comm_cost_multiagent} clearly shows that our WMAS achieves the best task results with the lowest token consumption, which validate its superior performance and minimal conversation overhead. Specifically, compared to GPTSwarm, WMAS achieves substantial reductions in token consumption of 31.4\%, 61.7\%, and 74.2\% on MMLU, GSM8K, and HumanEval, respectively. More specifically, WMAS reduces token consumption by approximately 17.5\% and 8.9\% compared to AutoGen and AgentPrune on MMLU, 39.5\% and 26.2\% on GSM8K, and 65.3\% and 64.5\% on HumanEval, respectively.
This advantage stems from our objective function and self-optimization process, which explicitly encourages sparsity in the conversation graph and thereby promotes more efficient interactions in WMAS.\par
\begin{figure}[t]
    \centering
    \includegraphics[width=0.95\linewidth]{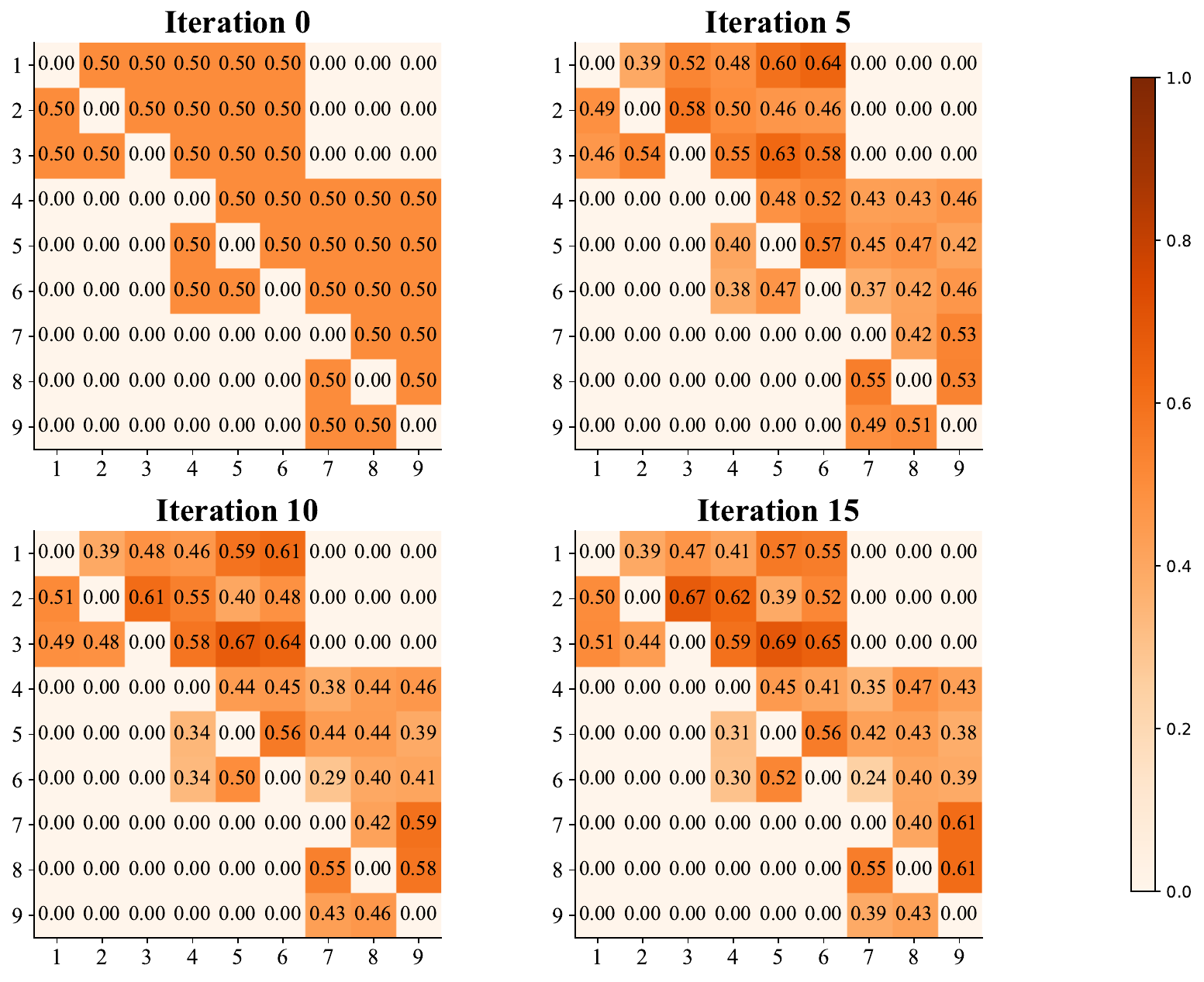}
    \caption{Evolution of the probabilistic heatmap $\mathbf{H}$ during the optimization process on the HumanEval dataset. }
    \label{fig:heatmap_evolution}
\end{figure}

Fig.~\ref{fig:heatmap_evolution} visualizes the evolution of the probabilistic heatmap $\mathbf{H}$ during optimization. As the optimization progresses, the probabilities associated with different directed edges begin to diverge significantly. Redundant edges that are less relevant to the task exhibit lower heat values, while edges that are more critical for task execution show higher intensities. 

\section{Conclusion}
In this paper, we propose a Wireless Multi-Agent System~(WMAS) towards intelligent and customized wireless networks. The WMAS can provide customized services for various user requests through multi-round conversations.
As the conversation topology plays an important role in MAS, we develop an optimization approach that allows the conversation topology to be self-optimized in terms of execution performance and conversation overhead. This approach is built on graph-based modeling, continuous relaxation, and reinforcement learning.
Extensive experiments on various tasks demonstrate that WMAS achieves superior task performance while significantly reducing conversation overhead compared to other MAS frameworks. 
These results validate the effectiveness of our WMAS to support effective and efficient task execution across a wide range of application scenarios.
In future work, WMAS can be utilized to provide users with constantly emerging new types of services, including AI, sensing, and so on, since WMAS has the ability to orchestrate MASs to tackle different types of tasks, and has the intelligence to improve its performance.

%%%%第二个表可以用图直接表示
% \begin{table}[htbp]
%     \centering
%     \caption{Communication cost of multi-agent systems on MMLU, GSM8K, and HumanEval. All values are measured in tokens.}
%     \label{tab:comm_cost_multiagent}
%     \begin{tabular}{l|c|c|c}
%         \hline
%         \textbf{Method} & \textbf{MMLU} & \textbf{GSM8K} & \textbf{HumanEval} \\
%         \hline
%         AutoGen       &    1,725,774       &      1,903,738
%           &          614,001         \\
%         GPTSwarm      &    2,075,918       &     3,007,838
%            &    920,215                 \\
%         AgentPrune    &   1,562,806          &  1,561,450           &    668,042                \\
%         \hline
%         \textbf{WMAS} &    1,424,535        &    1,152,071         &      237,416             \\
%         \hline
%     \end{tabular}
% \end{table}

% Evolution of the communication heatmap $\mathbf{H}$ during the optimization process. 
%     Each cell indicates the communication intensity from one agent to another, with higher values representing stronger interactions. 
%     The heatmaps illustrate the emergence of structured communication patterns over iterations.

\bibliographystyle{IEEEtran.bst}

\bibliography{refs_update}

\end{document}